\documentclass[12pt, a4paper]{article}
\usepackage[latin1]{inputenc}
\usepackage{vmargin}
\usepackage{amsfonts}
\usepackage{amssymb}
\usepackage{epstopdf}
\usepackage[dvips]{graphicx}
\usepackage{graphicx}
\usepackage{amsmath}
\usepackage{amsthm}
\usepackage{setspace}
\usepackage{multirow}
\usepackage{hyperref}
\usepackage{cases}
\usepackage[bottom]{footmisc}
\usepackage[small]{caption}
\newtheorem{Theorem}{Theorem}
\newtheorem{Proposition}{Proposition}
\newtheorem{Corollary}{Corollary}
\newtheorem{Lemma}{Lemma}
\begin{document}

\title{Permanent market impact can be nonlinear\thanks{This research has been conducted with the support of the Research Initiative ``Exécution optimale et statistiques de la liquidité haute fréquence'' under the aegis of the Europlace Institute of Finance. I would like to thank Robert Almgren (NYU and Quantitative Brokers), Jim Gatheral (CUNY Baruch), Nicolas Grandchamp des Raux (HSBC France), Charles-Albert Lehalle (Capital Fund Management), Guillaume Royer (Ecole Polytechnique) and Christopher Ulph (HSBC) for the conversations we had on market impact modeling. The points of view developed in this paper are solely those of the author.}}

\author{Olivier Gu\'eant\footnote{Universit\'e Paris-Diderot, UFR de Math\'ematiques, Laboratoire Jacques-Louis Lions. 175, rue du Chevaleret, 75013 Paris, France. \texttt{olivier.gueant@ann.jussieu.fr}}}
\date{}

\maketitle

\begin{center}
\textbf{Abstract}
\end{center}

There are two schools of thought regarding market impact modeling. On the one hand, seminal papers by Almgren and Chriss introduced a decomposition between a permanent market impact and a temporary (or instantaneous) market impact. This decomposition is used by most practitioners in execution models. On the other hand, recent research advocates for the use of a new modeling framework that goes down to the resilient dynamics of order books: transient market impact. One of the main criticisms against permanent market impact is that it has to be linear to avoid dynamic arbitrage. This important discovery made by Huberman and Stanzl \cite{huberman2004price} and Gatheral \cite{gatheral2010no} favors the transient market impact framework, as linear permanent market impact is at odds with reality. In this paper, we reconsider the point made by Gatheral using a simple model for market impact and show that permanent market impact can be nonlinear. Also, and this is the most important part from a practical point of view, we propose different statistics to estimate permanent market impact and execution costs that generalize the ones proposed in \cite{almgren2005direct}.

\section*{Introduction}

Over the course of the last decade, following the seminal papers on optimal liquidation by Almgren and Chriss \cite{almgren1999value, almgren2001optimal, almgren2003optimal, almgren2005direct}, a new strand of literature has emerged regarding the execution of large blocks of shares. Researchers, along with practitioners, have first adopted the modeling framework introduced by Almgren and Chriss, and developed or used variants of the initial models. At the heart of this framework, there is a decomposition of market impact into two separate parts. On the one hand, there is a permanent market impact, pushing prices up over the course of a buy order -- resp. pushing prices down over the course of a sell order --, hence modifying permanently the price process. On the other hand, there is a temporary market impact, sometimes called instantaneous market impact or simply phrased execution cost, paid by the trader to take liquidity from the market. Surprisingly, this framework has rarely been questioned until the development of a new literature that models the resilience of the order book using what is now called transient market impact. Research has long been focused indeed on the framework proposed by Almgren and Chriss. Generalization of Almgren and Chriss initial models have been proposed to account for stochastic volatility or stochastic liquidity (see \cite{almgren2011optimal}). Researchers also considered mean-variance and mean-quadratic variation optimization criteria (\cite{almgren2007adaptive}, \cite{forsyth2009optimal}, \cite{lorenz2010mean} and \cite{tse2011comparison}), along with CARA, IARA and DARA utility functions (see \cite{gueantnew}, \cite{schied2009risk}, \cite{schied2010optimal}). The first attempt to use a different model was in a working paper by Obizhaeva and Wang \cite{obizhaevaoptimal}.\footnote{This paper appeared for the first time on the Internet in 2005. It has been published far later in 2012.} For the first time, the authors decided to go from a statistical model of market impact (as the one proposed by Almgren and Chriss) to a descriptive model of order book resilience. This was the first model involving what is now called transient market impact. This second generation of optimal liquidation models, based on transient market impact, has developed in recent years (\cite{alfonsi2008constrained}, \cite{alfonsi2010optimal}, \cite{alfonsi2010optimal1} and \cite{predoiu2010optimal}). It raised the theoretical question of the functional forms for market impact that are compatible with the absence of price manipulation (see \cite{alfonsi2010optimal2}, \cite{alfonsi2009order}, \cite{gatheral2010no}, \cite{gatheral2010transient}, \cite{gatheral2011transient}, \cite{klock2011existence}). This question of the admissible forms for market impact had already been considered for permanent market impact by Huberman and Stanzl \cite{huberman2004price} but the most cited paper on the topic is the paper by Gatheral \cite{gatheral2010no} who states that permanent market impact has to be linear to avoid dynamic arbitrage, or in other words, to avoid round trips that are profitable on average.\\
No model using the Almgren-Chriss framework featured a nonlinear specification for the permanent market impact. However, the fact that it was apparently impossible turned out to be a problem. Indeed, in parallel to the study of market impact from a theoretical point of view, market impact estimates were carried out and they created a consensus toward a square root form for the impact -- only Almgren and coauthors fitted in \cite{almgren2005direct} a close-to-linear functional form for permanent market impact.\\
The goal of this paper is to reconciliate the Almgren-Chriss decomposition of market impact with reality, and to show that nonlinear permanent market impact is compatible with the absence of dynamic arbitrage. We indeed propose a framework in which permanent market impact is not only a function of the traded volume but also of the cumulated volume executed so far. This framework allows to have at the same time a nonlinear permanent market impact and the absence of dynamic arbitrage. We also show, under a power-law hypothesis for permanent market impact, that the cumulated instantaneous market impacts can be extracted from the slippage and both the pre-trade and post-trade prices, leading therefore to a generalization of the formula proposed by Almgren et \emph{al.} in \cite{almgren2005direct}.\\

Section 1 is dedicated to the classical Almgren-Chriss framework and it recalls the result of Gatheral. Section 2 presents the main idea of this paper to model permanent market impact. It shows that the modeling framework is compatible with both a nonlinear permanent market impact and the absence of dynamic arbitrage. Section 3 shows how the cumulated instantaneous market impacts can be extracted from slippage, pre-trade price and post-trade price. This point is important for statistical purposes. It generalizes \cite{almgren2005direct}.

\section{Almgren-Chriss framework and Gatheral's result}

\subsection{Framework}

Let us fix a probability space $(\Omega, \mathcal{F}, \mathbb{P})$ equipped with a filtration $(\mathcal{F}_t)_{t\in \mathbb{R}_+}$ satisfying
the usual conditions. We assume that all stochastic processes are defined on $(\Omega, \mathcal{F},(\mathcal{F}_t)_{t\in \mathbb{R}_+}, \mathbb{P})$.\\

We consider a trader who wants, over a time window $[0,T],$ $T>0$, to go from a portfolio made of $q_0$ shares of a given stock to a portfolio made of $q_T$ shares of the same stock. For that purpose, we consider absolutely continuous strategies $q : [0,T] \to \mathbb{R}$ with $q(0) = q_0$ and $q(T) = q_T$.\\

The execution process usually impacts stock price. The usual framework to model market impact is the following. First, the price $S$ of the stock has the following dynamics:

$$dS_t = \sigma dW_t - k(v_t) dt,$$ where $v_t = -q'(t)$, and where $k(v)$ has the same sign as $v$: a buy order pushes the price up, while a sell order pushes the price down.\\

Besides this permanent market impact on price, the price obtained for a trade at time $t$ is not $S_t$ but rather $S_t - h\left(t,v_t\right)$ where $h(t,v)$ is assumed to have the same sign as $v$.\\
The resulting dynamics for the cash account $X$ is:

$$dX_t = v_t S_t dt - h\left(t,v_t\right) v_t dt.$$

\subsection{Gatheral's result}

The first result we state is the value of the cash account at the end of the trading period: $X_T$.

\begin{Proposition}
\label{X}
$$X_T = X_0 + (q_0-q_T) S_0 + \int_0^T (q_T-q_t) k(v_t) dt  - \int_0^T h\left(t,v_t\right) v_t dt  - \int_0^T \sigma (q_T-q_t) dW_t.$$
\end{Proposition}

\emph{Proof:}\\

We write the definition of $X_T$ and we integrate by parts to obtain:
\begin{eqnarray*}
  X_T &=& X_0+ \int_0^T v_t S_t dt - \int_0^T h(t,v_t) v_t dt
\end{eqnarray*}
\begin{eqnarray*}
   &=& X_0 + \left[(q_T-q_t)S_t\right]^T_0  +  \int_0^T (q_T-q_t) k(v_t)  dt\\
    &&- \int_0^T \sigma (q_T-q_t) dW_t - \int_0^T h\left(t,v_t\right) v_t dt \\
   &=& X_0 + (q_0-q_T) S_0 + \int_0^T (q_T-q_t) k(v_t) dt\\
   &&  - \int_0^T h\left(t,v_t\right) v_t dt  - \int_0^T \sigma (q_T-q_t) dW_t.
\end{eqnarray*}\qed\\

Then, following Gatheral (\cite{gatheral2010no}), we define a dynamic arbitrage as a round trip from $q_0$ to $q_T=q_0$ that is profitable on average. In other words, there is a dynamic arbitrage if and only if there exists an absolutely continuous function $q : [0,T] \to \mathbb{R}$ such that the two following assertions are true:

\begin{itemize}
  \item $q(0) = q(T)$,
  \item The associated cash process verifies $\mathbb{E}[X_T] \ge X_0$.
\end{itemize}

Gatheral shows in \cite{gatheral2010no} that a necessary and sufficient condition for the absence of dynamic arbitrage is that $k$ is a linear function. More precisely:

\begin{Theorem}[Gatheral's Theorem]
\label{gath}
We have:
\begin{itemize}
  \item If $k(v) = k v$, then for any absolutely continuous function $q : [0,T] \to \mathbb{R}$, with $q(0) = q(T)$, the associated cash process verifies $\mathbb{E}[X_T] \le X_0$.\\
  \item If $h=0$ and if $k$ is not linear, there exists a dynamic arbitrage.\\
\end{itemize}

\end{Theorem}

Considering the case $h=0$ is a conservative assumption. The linearity assumption for $k$ ensures in fact that there is no dynamic arbitrage independently of $h$.\\

This result has very important consequences. The most important one concerns the overall price impact of a transaction when $k(v) = kv$. If we consider indeed a sell order of $q_0$ shares over a time window $[0,T]$, then, for any $T' \ge T$, the stock price at time $T'$ is given by:
$$S_{T'} - S_0 = -k q_0 + \sigma W_{T'}.$$

This means that the permanent market impact at the macroscopic level is linear, at odds with the square root (or at least strictly concave) shape usually obtained in the literature.\\
The goal of the next section is to recover a nonlinear and concave permanent market impact at the macroscopic level while avoiding the existence of dynamic arbitrage.

\section{Nonlinearity and no-dynamic arbitrage}

One of the limitations of the preceding framework is that the permanent market impact function at the microscopic level ($k$) was a function of $v$ only. The concavity of price impact observed empirically at the macroscopic level can in fact be interpreted in the following way. When a trading process starts, it adds new volume to the market and market participants anticipate that there will be more volume in the near future. Hence, they add volume to the market, decreasing therefore the impact of future trades. Mathematically, it means that we can replace the expression $k(v_t)$ by an expression of the form $f(|q_0-q_t|)v_t$, where $f$ is a positive and decreasing function: the impact is still linear at the microscopic level but it decreases as the quantity executed so far is increasing.\\

The above remarks lead to the following model:
$$dq_t = -v_t dt,$$
$$dS_t = \sigma dW_t - f\left(|q_0-q_t|\right) v_tdt,$$
$$dX_t = v_t S_t dt - h(t,v_t) v_t dt,$$
where $f$ is assumed to be a positive function in $L^1_{loc}(\mathbb{R}_+)$.\\

Our first result concerns the value of the cash process at the end of the execution process:

\begin{Lemma}
\label{X2}
$$X_T = X_0 + (q_0-q_T) S_0 + \int_0^T (q_T-q_t) f(|q_0-q_t|)v_t dt  - \int_0^T h\left(t,v_t\right) v_t dt  - \int_0^T \sigma (q_T-q_t) dW_t.$$
\end{Lemma}

\emph{Proof:}\\

The proof is the same as above:
\begin{eqnarray*}
  X_T &=& X_0+ \int_0^T v_t S_t dt - \int_0^T h(t,v_t) v_t dt\\
   &=& X_0 + \left[(q_T-q_t)S_t\right]^T_0  +  \int_0^T (q_T-q_t) f(|q_0-q_t|) v_t dt\\
    &&- \int_0^T \sigma (q_T-q_t) dW_t - \int_0^T h\left(t,v_t\right) v_t dt \\
   &=& X_0 + (q_0-q_T) S_0 + \int_0^T (q_T-q_t) f(|q_0-q_t|) v_t dt\\
   &&  - \int_0^T h\left(t,v_t\right) v_t dt  - \int_0^T \sigma (q_T-q_t) dW_t.
\end{eqnarray*}\qed\\

Now, an important point of our paper is that there is no dynamic arbitrage in our model. More precisely:

\begin{Proposition}[No-dynamic arbitrage]
\label{nda}
For any absolutely continuous function $q : [0,T] \to \mathbb{R}$, with $q(0) = q(T)$, the associated cash process verifies $\mathbb{E}[X_T] \le X_0$.\\
\end{Proposition}

\emph{Proof:}\\

Using Lemma \ref{X2} we have:

$$X_T = X_0 + \int_0^T (q_0-q_t) f(|q_0-q_t|)v_t dt  - \int_0^T h\left(t,v_t\right) v_t dt  - \int_0^T \sigma (q_0-q_t) dW_t.$$

Let us define $G(z) = \int_0^z y f(|y|)dy$. The above expression can be written as:

\begin{eqnarray*}
  X_T &=& X_0 + \int_0^T G'(q_0-q_t) v_t dt  - \int_0^T h\left(t,v_t\right) v_t dt  - \int_0^T \sigma (q_0-q_t) dW_t\\
   &=& X_0 + [G(q_0-q_t)]_0^T - \int_0^T h\left(t,v_t\right) v_t dt  - \int_0^T \sigma (q_0-q_t) dW_t \\
   &=& X_0 - \int_0^T h\left(t,v_t\right) v_t dt  - \int_0^T \sigma (q_0-q_t) dW_t.
\end{eqnarray*}

Hence:

$$\mathbb{E}[X_T] = X_0 - \int_0^T h\left(t,v_t\right) v_t dt \le X_0.$$\qed

In the model we propose, there is no-dynamic arbitrage. However, it does not lead to a linear macroscopic permanent market impact. If one considers indeed a trade of size $q_0$ over the time window $[0,T]$, then the price move due to the transaction is characterized by the following Proposition:

\begin{Proposition}[Nonlinear permanent market impact]
\label{nonlin}
Let us define $F(z) = \int_0^z f(|y|) dy$.\\
If $q(0)=q_0$ and $q(T) = 0$, then $\forall T' \ge T$:
$$S_{T'} - S_0 = -F(q_0) + \sigma W_{T'}.$$
\end{Proposition}

\emph{Proof:}\\

\begin{eqnarray*}
  S_{T'} - S_0 &=& \sigma  W_{T'} - \int_0^{T} v_t f\left(|q_0-q_t|\right)dt\\
   &=&  \sigma W_{T'} - [F(q_0-q_t)]_0^T\\
   &=& \sigma W_{T'} - F(q_0).
\end{eqnarray*}
\qed\\

Nonlinear permanent market impact is then compatible with no dynamic arbitrage. In particular, most estimations of permanent market impact lead to a function $F$ of the form $F: q \mapsto k sgn(q)|q|^\alpha$, with $k>0$ and $\alpha \simeq 0.5$.\\
This means that $f$ is of the form $f: q \in \mathbb{R}^*_+ \mapsto \frac{k\alpha}{q^{1-\alpha}}$. In particular, when $\alpha \in (0,1)$, $f$ blows up at $0$. This is not a real problem since the function remains in $L^1_{loc}(\mathbb{R}_+)$. Obviously, one can replace $\frac{k\alpha}{q^{1-\alpha}}$ by $\frac{k\alpha}{(q+A)^{1-\alpha}}$ where $A>0$ to prevent the blow up.\\

\section{Optimal execution models and estimation}

We have shown that our model has two characteristics that were supposed to be incompatible: no dynamic arbitrage and nonlinear permanent market impact. Also, the model can be used in optimal execution models since the effect of permanent market impact is independent of the liquidation strategy. This can be seen in the following Lemma that gives the expression of the cash account at the end of a liquidation process.

\begin{Lemma}
\label{X3}
Let us consider the special case $q_T=0$. Let us define $F(z) = \int_0^z f(|y|)dy$. Then:
$$X_T= X_0 + q_0 S_0 - \int_0^{q_0} F(z) dz - \int_0^T h\left(t,v_t\right) v_t dt  + \int_0^T \sigma q_t dW_t.$$
In particular, the influence of the permanent market impact on $X_T$ is independent of the strategy $q$ and only depends on $q_0$.\\
\end{Lemma}

\emph{Proof:}\\

We use Lemma \ref{X2} when $q_T=0$ to obtain:

\begin{eqnarray*}
  X_T &=& X_0 + q_0 S_0 + \int_0^T -q_t f(|q_0-q_t|)v_t dt  - \int_0^T h\left(t,v_t\right) v_t dt  + \int_0^T \sigma q_t dW_t\\
   &=& X_0 + q_0 S_0 + [-q_t F(q_0-q_t)]_0^T - \int_0^T v_t F(q_0-q_t) dt\\
   &&- \int_0^T h\left(t,v_t\right) v_t dt  + \int_0^T \sigma q_t dW_t\\
  &=&  X_0 + q_0 S_0 - \int_0^{q_0} F(y) dy - \int_0^T h\left(t,v_t\right) v_t dt  + \int_0^T \sigma q_t dW_t
\end{eqnarray*}
\qed\\

If one considers the special case of a power function for permanent market impact, then we obtain the following straightforward corollary:

\begin{Corollary}
\label{c1}
If $f: q \in \mathbb{R}^*_+ \mapsto \frac{k\alpha}{q^{1-\alpha}}$ with $\alpha \in (0,1]$ and $q_T=0$ then:
$$X_T= X_0 + q_0 S_0 - \frac{k}{1+\alpha} |q_0|^{1+\alpha} - \int_0^T h\left(t,v_t\right) v_t dt  + \int_0^T \sigma q_t dW_t.$$
\end{Corollary}

This expression will allow to recover the cumulated instantaneous market impacts $\int_0^T h\left(t,v_t\right) v_t dt$:\\

\begin{Proposition}
Under the assumptions of Corollary \ref{c1}, we have:
$$ \frac{S_{T'} + \alpha S_0}{1+\alpha} - \frac{X_T - X_0}{q_0} = \frac{1}{q_0}\int_0^T h\left(t,v_t\right) v_t dt - \sigma \int_0^T \frac{q_t}{q_0} dW_t + \frac{\sigma}{1+\alpha} W_{T'}$$
\end{Proposition}

\emph{Proof:}\\

We have:
$$X_T - X_0 - q_0 S_0 = -\frac{k}{1+\alpha} |q_0|^{1+\alpha} - \int_0^T h\left(t,v_t\right) v_t dt  + \int_0^T \sigma q_t dW_t$$
and
$$S_{T'} - S_0 = \sigma W_{T'} - k sgn(q_0)|q_0|^\alpha.$$

Hence:

$$X_T - X_0 - q_0 S_0 = -\frac{q_0}{1+\alpha} \left(\sigma W_{T'} - (S_{T'} - S_0) \right) - \int_0^T h\left(t,v_t\right) v_t dt  + \int_0^T \sigma q_t dW_t.$$

Reorganizing the terms, we get:
$$ \frac{S_{T'} + \alpha S_0}{1+\alpha} - \frac{X_T - X_0}{q_0} = \frac{1}{q_0}\int_0^T h\left(t,v_t\right) v_t dt - \frac{\sigma}{S_0} \int_0^T \frac{q_t}{q_0} dW_t + \frac{\sigma}{1+\alpha} W_{T'}.$$\qed\\

This Proposition is important because it can be rewritten:

$$ \frac{q_0 S_0 - (X_T - X_0)}{q_0S_0} + \frac{1}{1+\alpha} \frac{S_{T'} - S_0}{S_0}$$$$ = \frac{1}{q_0 S_0}\int_0^T h\left(t,v_t\right) v_t dt - \sigma \int_0^T \frac{q_t}{q_0} dW_t + \frac{\sigma}{(1+\alpha)S_0} W_{T'},$$
or equivalently:\footnote{The quantities below are signed: slippage and cumulated instantaneous market impact are positive when $q_0$ is positive (sell order) and negative when $q_0$ is negative (buy order).}

$$\textrm{Slippage }(\%) + \frac{1}{1+\alpha} \textrm{Price\; Return} (\%) = \textrm{Cum.\; Inst.\; Market\; Impact} (\%) + \textrm{Noise}.$$

Hence, both permanent market impact and instantaneous market impact can be estimated using three variables: (i) slippage, (ii) pre-trade price and (iii) post-trade price. This is more clearly stated in the following Theorem that generalizes the result of Almgren et \emph{al.} (\cite{almgren2005direct}) -- the case of Almgren et \emph{al.} corresponding to $\alpha=1$:

\begin{Theorem}
Under the assumptions of Corollary \ref{c1} and denoting $T' = T + \delta$, we have:
$$ S_{T'} - S_0 = - k sgn(q_0)|q_0|^\alpha + \epsilon_1$$
$$ \frac{S_{T'} + \alpha S_0}{1+\alpha} - \frac{X_T - X_0}{q_0} = \frac{1}{q_0}\int_0^T h\left(t,v_t\right) v_t dt + \epsilon_2$$
where
$$(\epsilon_1,\epsilon_2) = \mathcal{N}\left(0,\sigma^2 \left(
                                                                        \begin{array}{cc}
                                                                          T+ \delta & \frac{\delta}{1+\alpha} - \int_0^T \left(\frac{q_t}{q_0} - \frac{1}{1+\alpha}\right) dt \\
                                                                          \frac{\delta}{1+\alpha} - \int_0^T \left(\frac{q_t}{q_0} - \frac{1}{1+\alpha}\right) dt & \frac{\delta}{(1+\alpha)^2} + \int_0^T \left( \frac {q_t}{q_0} - \frac {1}{1+\alpha} \right)^2 dt\\
                                                                        \end{array}
                                                                   \right)\right).$$

In particular, if we assume that $q_t = q_0 \left(1-\frac tT\right)$, then:
$$(\epsilon_1,\epsilon_2) = \mathcal{N}\left(0,\sigma^2 \left(
                                                                        \begin{array}{cc}
                                                                          T+ \delta & \frac{\delta}{1+\alpha} + T\left(\frac{1}{1+\alpha} - \frac 12\right) \\
                                                                          \frac{\delta}{1+\alpha} + T\left(\frac{1}{1+\alpha} - \frac 12\right) & \frac{\delta}{(1+\alpha)^2} + \frac T 3 \frac{1+\alpha^3}{(1+\alpha)^3}\\
                                                                        \end{array}
                                                                   \right)\right).$$
\end{Theorem}

\emph{Proof:}\\

$$\epsilon_1 = \sigma W_{T'}, \qquad \epsilon_2 = - \sigma \int_0^T \frac{q_t}{q_0} dW_t + \frac{\sigma}{1+\alpha} W_{T'}.$$

We can compute the variance of $\epsilon_1$ and $\epsilon_2$ and the associated covariance:
$$\mathbb{V}(\epsilon_1) = \sigma^2 T'$$
$$\mathbb{V}\left(\epsilon_2\right) = \sigma^2 \left(\frac{\delta}{(1+\alpha)^2} + \int_0^T \left( \frac {q_t}{q_0} - \frac {1}{1+\alpha} \right)^2 dt \right)$$
$$Cov(\epsilon_1,\epsilon_2) = \sigma^2 \left(\frac{\delta}{1+\alpha} - \int_0^T \left(\frac{q_t}{q_0} - \frac{1}{1+\alpha}\right) dt \right).$$
The particular case where $q_t = q_0 \left(1-\frac tT\right)$ is then obtained straightforwardly.\qed\\

This Theorem generalizes a result by Almgren and coauthors (in the case $\alpha=1$). This result was used to estimate the two components of market impact using a database of metaorders (assumed to be executed with a POV algorithm) and their decomposition into child orders. As permanent market impact is not linear, one needs to use the above Theorem to estimate the parameters using Almgren's estimation methodology. Otherwise, the estimation of the instantaneous market impact function is biased by a term coming from permanent market impact.\\

\section*{Conclusion}

In this paper, we propose a new modeling framework compatible with both no dynamic arbitrage and nonlinear permanent market impact at the macroscopic level. This framework is rooted into the classical decomposition between permanent and instantaneous market impacts. It allows to replicate the very robust stylized fact of a concave permanent market impact, in particular the square root shape usually observed. Interestingly, it does not change anything to classical optimal execution models since the permanent component of market impact continues to have no influence on optimal strategies. However, the classical estimation of the instantaneous market impact function needs to be reconsidered and we propose in Theorem 2 new equations that generalize \cite{almgren2005direct} to estimate market impact functions.\\

\end{document}